\documentclass[12pt,preprint]{aastex}

\slugcomment{Article for The Astronomical Journal}

\shorttitle{Hydro-Gravitational Theory and Observations}
\shortauthors{Gibson and Schild}

\begin{document}


\title{Evidence for Hydro-Gravitational Structure Formation Theory versus 
Cold-Dark-Matter, Hierarchical-Clustering, and Jeans 1902}


\author{Carl H. Gibson\altaffilmark{1}}
\affil{Departments of Mechanical and Aerospace Engineering and  Scripps
Institution of Oceanography, University of California,
      San Diego, CA 92093-0411}

\email{cgibson@ucsd.edu}

\and

\author{Rudolph E. Schild}
\affil{Center for Astrophysics,
      60 Garden Street, Cambridge, MA 02138}
\email{rschild@cfa.harvard.edu}


\altaffiltext{1}{Center for Astrophysics and Space Sciences, UCSD}


\begin{abstract} Observations are compared to conflicting predictions about
self-gravitational structure formation by the hydro-gravitational theory
(HGT) of Gibson 1996-2003 versus cold-dark-matter
hierarchical-clustering-cosmology (CDMHCC) and the Jeans 1902 criterion.
According to HGT, gravitational structures form immediately after mass-energy
equality by plasma fragmentation at 30,000 years when viscous and weak turbulence
forces first balance gravitational forces within the horizon
$L_H \equiv ct < L_J = c/[3\rho G]^{1/2}$,
contrary to the Jeans 1902 criterion.  Buoyancy forces fossilize the
$10^{-12}$ s$^{-1}$ rate-of-strain and the $10^{-17}$ kg m$^{-3}$ baryonic
density. The non-baryonic dark matter (NBDM) diffuses into the voids
rather than forming cold-dark-matter (CDM) halos required by
CDMHCC.  From HGT, supercluster-mass  to galaxy-mass fragments exist at the
plasma to gas transition, and these fragment further to form proto-globular-star
clusters (PGCs) and planetary-mass primordial-fog-particles (PFPs): the
baryonic dark matter of the interstellar-medium and
inner-galaxy-dark-matter-halos, from which all planets and stars are formed by
accretion (Gibson 1996, Schild 1996).  From HGT and a rich cluster
mass profile \citep{tys95}, $D_{NBDM}\approx 6 \times
10^{28}$ m$^2$ s$^{-1}$, 
$m_{NBDM}
\le 10^{-33}$ kg, and the NBDM forms outer-galaxy halos after 300,000 years.
\end{abstract}




\section{Introduction} Observations are accumulating about the
early universe and the interstellar medium that challenge the standard models of
cosmology, galaxy formation, star formation, planet formation and comet
formation.  For many years no data were available to contradict the Jeans 1902
gravitational structure formation criterion that in a gas of uniform density
$\rho$, gravitational condensation cannot occur on scales smaller than
$L_J$ where
\begin{equation}  L_J \equiv V_S/(\rho G)^{1/2}
\gg (p/\rho^2 G)^{1/2} \approx (RT/\rho G)^{1/2} \equiv L_{JHS} ,
\label{eq1}
\end{equation}
   $G$ is Newton's gravitational constant, $R = k/m$ is the gas constant, $k$
is Boltzmann's constant, $m$ is the gas particle mass, and $ T$ is the
temperature.  Because 
$V_S \approx (p/\rho)^{1/2} \approx (RT)^{1/2}$ is $V_S$ for an
ideal gas,  Jeans 1902 and others have assumed that the Jeans acoustic scale
$L_J$ defined on the left and the Jeans hydrostatic scale
$L_{JHS}$ defined on the right of (\ref{eq1}) are equivalent. They are not.  It
has been shown that the Jeans criterion is unreliable \citep{gib96}: non-acoustic
density fluctuations are absolutely unstable to structure formation. Hydrostatic
equilibrium is achieved only at small scales $L_{JHS}$ determined by the largest
viscous, turbulent, or diffusive Schwarz scale (Table 1) of HGT and various
planetary and star formation processes that are induced at positive non-acoustic
nuclei, with $L_J \gg L_{JHS}$.   Using viscous forces, turbulent forces and
diffusion it was shown from HGT-cosmology that the baryonic dark matter is
dominated by small frozen planets ($10^{-6} M_{\sun}$ primordial fog particles,
or PFPs) that fragmented by self gravity at the plasma to gas transition 300,000
years after the big bang, $\S 2$.  This theoretical prediction was immediately
and independently supported  by observations of twinkling frequencies of a lensed
quasar, suggesting the lensing galaxy mass is dominated by rogue planets
``likely to be the missing mass''
\citep{sch96}.   The Gibson $\& $ Schild 1996 prediction and observational
demonstration that the baryonic dark matter is primordial
micro-brown-dwarfs has since been confirmed by independent observations in the
same and other lensed quasars, and by other observational evidence,
$\S3$.

HGT describes
the formation of structure by self-gravitational condensations and rarefactions
for cosmology, astrophysics and astronomy.  The first gravitational structures
to form were proto-supercluster-voids in the plasma epoch by fragmentation at
30,000 years after the big bang when viscous and weak turbulent forces first
matched gravity forces at the horizon scale $L_H$, followed by
proto-cluster-voids and proto-galaxy-voids filled with the weakly collisional
non-baryonic-dark-matter (NBDM) by diffusion \citep{gib00}.  Cold dark matter
condensations (CDM-halos) are excluded by HGT for scales smaller than $L_{SD}
\gg L_H$ in the plasma epoch, and are not necessary to make structure in the gas
epoch (after 300,000 years) by hierarchical-clustering-cosmology (HCC), which is
also excluded by HGT.  Gravitational clustering of collisionless CDM-halos would
produce steep core cusps in galaxies that are not observed
\citep{sa02}.  The most distant galaxies observed are clustered \citep{sti00}
and the closest clusters observed are fragmenting \citep{gs03c}, contrary to HCC.
The early clusters predicted by HGT serve as gravitational lenses and produce 
arcs and multiple arcs of background galaxies at  high rates of incidence that
contradict all the flat-CDM cosmologies by orders of magnitude
\citep{bar98}. It is necessary to invent special ``super-lensing'' clusters to
explain the multiple arcs observed in the recent Red-Sequence Cluster Survey and
other studies, which would otherwise exclude the presently popular
flat-$\Lambda$CDM cosmology by orders of magnitude
\citep{gla03}. Drag forces from the expansion of the universe
rapidly separated proto-galaxies and helped fragment proto-galaxy-clusters at the
plasma-gas transition  \citep{gib99a}, making galaxy mergers quite rare rather
than mandatory as in HCC, and explaining alignments of bright central
supercluster galaxies with background QSOs by perspective that otherwise would
confute CDMHCC, the big bang, and standard physics  \citep{hoy00} without HGT
\citep{gs03a}.   Two of these rare galaxy mergers were imaged by the new HST/ACS
wide angle camera in the Tadpole complex VV29abcdef \citep{gs03b} and in Mice,
showing star trails of young-globular-clusters (YGCs) produced as star formation
is triggered from PGC-PFPs by the mergers of galaxies and their
baryonic-dark-matter (BDM) halos.  High resolution wide angle images by HST/ACS
in the Helix planetary nebula and other nearby PNe in the Galaxy support the HGT
prediction that galaxy disks are formed by accreted PGCs, and that the ISM has
the high primordial PGC-PFP density fossilized at the time of first structure
and expected in star forming regions
\citep{gs03c}.

From HGT, density perturbations $\delta \rho (t)$ on scales $L < L_J$ in a
uniform density gas are absolutely unstable to the formation of gravitational
structure, and grow or decrease exponentially with time depending on the
sign of the perturbation (see Reyden 2003 $\S 12.1$ but ignore $\S 12.2$). Let
$\rho(t) =
\rho + \delta \rho (t)$ inside a sphere of radius $\vec{r}$ and constant
$\rho$ outside, where $\delta \rho (0) = \delta \rho(t=0)$ and $L_C \ll
|\vec{r}|
\ll L_J$ with
$L_C$ the particle collision length.  As correctly shown in
$\S 12.1$, self gravitation causes the density perturbation to either increase
in magnitude and shrink in size (gravitational condensation, $\delta\rho(t) =
\delta\rho(0) exp (t \surd [4 \pi G\rho] ) $, with $ \delta\rho(0) > 0$), or
increase in magnitude and grow in size (gravitational void formation,
$\delta\rho(t) = \delta \rho(0) exp (-t \surd [4 \pi G \rho] ) $, with $
\delta\rho(0) < 0$).  Non-acoustic density fluctuations are absolutely
unstable to self-gravitational formation of structure, just as shear layers are absolutely unstable
to the formation of turbulence due to
inertial-vortex forces $\vec{v} \times \vec{\omega}$, where $\vec{v}$ is velocity and $\vec{\omega}$
is vorticity \citep{gib99b}.  Both processes are highly non-linear so neither can
be reliably described by linear perturbation stability analysis, and both couple
very large scales to very small scales so they are difficult to simulate
numerically.  It is stated without proof in
$\S 12.2$
\citep{rey03} that pressure will prevent the exponential growth of density
perturbations on length scales
$L
\ll L_J$.  This is not true.  The idea that ``pressure support'' or ``thermal
support'' prevents structure formation on scales smaller than $L_J$ is an
unfortunate legacy of the erroneous Jeans 1902 criterion for structure
formation.

The time
$t_{S}$ required for pressure waves (sound) to move a distance
$L$ is
$t_{S} = L/V_S$, where $V_S$ is the speed of sound.  For
$t_{S}=t_{FF}\equiv ( G
\rho)^{-1/2}$,
$L = L_J$, so $\bigtriangledown p
\rightarrow 0$  for $|\vec{r}| \ll L_J$ and $t_{S} \le t \le t_{FF}
$. Density perturbations $\delta
\rho(r\le r_0,t=0)$ of size $r_0$ grow to large values for $t \ge
t_{FF}$ as we have seen.   Pressure forces 
have the wrong sign to resist gravitational condensation except near the
gravitational stagnation point because they arise from Bernoulli's equation,
where
$p/\rho + v^2 /2
\approx constant$. Pressure forces overcome
condensational self  gravitational forces $ G \rho \delta \rho  L
$  on scales
$L_{JHS} < r_0
\ll L_J$ only at late stages in a collapse ($\delta
\rho > 0$)  due to local
irreversible processes (eg: turbulence stresses, viscous heating,
condensation of the gas to liquid form, freezing of the liquid, star
formation), and have no effect on small
scale self gravitational void formation forces ($\delta
\rho < 0$) until the void size reaches $L_J$.  The maximum speed of
void growth is the maximum speed of a rarefaction wave, which is the speed of
sound.  Self-gravity has the effect of a negative density diffusivity
$D_{\rho G}(L)
\approx - L^2  (\rho G)^{1/2}$ in the vicinity $L \le L_{SD}$ (see Table 1 and
$\S2$) of a non-acoustic density maximum or density minimum, with the resulting
absolute instability.

High resolution numerical simulations of stagnant gas show fragmentation and
condensation at scales smaller than the Jeans scale, but in calculations to
date these have been systematically filtered out as numerical artifacts
\citep{tru97}. The authors assume that pressure support or thermal
support prevent sub-Jeans-scale self-gravitational condensations and
fragmentations in astrophysical fluids and adjust their numerical simulations to
conform to these assumptions.

The large
gassy Jovian planets beyond Mars in the solar system are a mystery from the Jeans
criterion because they violate the Jeans limit
$M_J \ge 0.1 M_{\sun}$
\citep{lar85}.  If we compute the Jeans mass $M_J = L_J ^3 \rho $ and solve for
the Jeans temperature  $T_J$ we find
\begin{equation} T_J \approx M_J ^{2/3} \rho ^ {1/3} G R^{-1},
\end{equation} where $R = 8314 / M_{mol}$ m$^2$ s$^{-2}$ K$^{-1}$ is the gas
constant and
$M_{mol}$ is the gas molecular weight.  Substituting a  primordial H-He gas
constant
$R_{PM} \approx 3612$ m$^{2}$ s$^{-2}$K$^{-1}$ and the primordial (first
structure) gas density
$\rho = 10^{-17}$ kg m$^{-3}$ for Jupiter mass $\approx 10^{27}$ kg with $G= 6.7
\times 10^{-11}$ m$^{3}$ kg$^{-1}$ s$^{-2}$ gives a Jeans temperature $T_J
\approx 3.9 \times 10^{-2}$ K that is impossibly cold.   Why are Jupiter, Saturn, Uranus and Neptune all contrary to
the Jeans limit?  Numerous extra-solar planets with Jupiter mass have now been
detected that must also be composed  of H-He gas.  How are these possible?
How are rocky planets possible?  How are rocks possible?  All are impossible
by the Jeans criterion, and all are easily explained by HGT,
$\S2$.  
According to HGT, all large planets are formed by a hierarchical
clustering of PFPs to form Jovian-PFP-Planets (JPPs) and all stars are formed by
further accretion of PFP-JPPs and from their fragments and gas formed by the
close proximity to the proto-star.  Dust from supernovas is collected by
numerous PFP-JPPs in the ISM and turned into rocks near their cores and the
cores of the numerous large planets and brown dwarfs (JPPs) formed as the PFPs
accrete. Multiple hydrogen freeze-thaw cycles of wandering PFP-JPPs produces
many crushing central pressure episodes and layerings.
Rocky planets like Earth are thus rocky cores of Jupiters evaporated as the star
grows by JPP accretion.  

Long period comets with
random Oort cloud orbits $\approx 10^{16}$ m matching a typical PNe radius
suggest that the trillions of Oort comets are from a few hundred PFPs constantly
drizzling in from the sun's accretion hole in the ISM (its Oort cloud inner
radius according to HGT) that were not captured by the sun and its Jovians in
their first accretional attempt and have not yet been completely fragmented by
tidal forces and evaporated by radiation
\citep{gs03c}.  Orbit perturbations of 82 new (first-entry) class I (accurate)
Oort comets imply the existence of a
 SIRTF-infrared and VLA-radio detectable
$ 10^{28}$ kg JPP at $ 4 \times 10^{15}$ m 
 \citep{mat99}.
  From HGT the inner radius
of the Oort cloud is
$ 3
\times 10^{15}$ m, corresponding to a solar mass accreted from a PGC,
and the outer radius is
$ 3
\times 10^{17}$ m, corresponding to the average extent of a PGC.  Without HGT the
origin of the huge numbers of fragile, frozen,  Oort comets with their eccentric
orbits, specific distance scale, and random directions is mysterious.

What about the inter-stellar-medium (ISM)?  If there are really 30,000,000 PFPs
per star in the average galaxy, shouldn't some of these be revealed by their
evaporated gas near very hot objects; for example, near carbon white dwarfs
formed when ordinary stars have burned most of their H-He gas to form planetary
nebulae, or in the ISM after novas and supernovas?  Evidence that mysterious
``free-floating planets pepper our galaxy'' (June 2003 Astronomy cover page) is
rapidly accumulating
\citep{nae03}. The number of rogue Jupiters greatly exceeds predictions of
``failed brown dwarf'' and star expulsion models.  Why do
mysterious knots,  comets and Herbig-Haro objects appear from the dark
ISM whenever young stars form plasma jets?  Where does the huge amount of
gas in the halos of planetary nebulae come from?  Are ordinary stars really
so inefficient that most of their mass is wasted forming PNe halos?  See
$\S3$ for a discussion of HST images of the Helix Planetary Nebula \citep{gs03c},
which is the closest PNe to earth, and its thousands of cometary globule PFP and
JPP candidates.

HGT contradicts cold-dark-matter hierarchical-clustering-cosmologies
(CDMHCC).  CDM ``halos'' cannot remain gravitationally bound because the weakly
collisional CDM particles are intrinsically much too diffusive, with
$(L_{SD})_{CDM} \gg L_H$ during the plasma epoch.  Hence, small primordial CDM
``halos'' (the key misconception) cannot cluster to provide massive gravitational
potential wells to collect the baryonic matter and hierarchically form galaxies,
clusters and superclusters with ever increasing mass as proposed by CDMHCC. 
Instead, HGT forms protosuperclusters first in the hot plasma epoch and these
gently fragment to proto-clusters and then proto-galaxies as the plasma cools,
starting at 30,000 years and finishing at the plasma to gas transition at
300,000 years.  The gentle fragmentation from viscous-gravitational beginnings
in a uniformly expanding universe explains the remarkable angular and geometric
correlations observed between galaxies and quasi-stellar-objects (QSOs) with
highly discordant red shifts that are used to contradict the big bang hypothesis
\citep{hoy00}.

Compact galaxy clusters like Stephan's Quintet (SQ) with highly discordant
redshifts support the hypothesis that redshifts may be intrinsic
\citep{arp73} and that all the SQ galaxies are at the same distance and were
ejected by galaxy NGC 7331 located at the same redshift $z=0.0027$ as the
nearest member NGC 7320 of the SQ cluster.  No physical explanation has been
put forth for intrinsic redshifts and QSO ejections.  The observations
supporting these hypotheses can be easily explained by HGT without inventing
new physics by recognizing that the galaxies and QSOs are aligned by
perspective, and stay aligned along narrow lines of sight as the universe
expands because their sticky beginnings allow only small transverse
velocities.  CDMHCCs are contradicted by the accumulation of evidence showing
strong angular coincidences of QSOs and bright galaxies
\citep{bur03}. Stephan's Quintet is presented as an example in $\S3$.

\section{Hydro-Gravitational Theory}

Standard CDM cosmologies are flawed by their adoption of simplistic fluid
mechanical equations that fail to incorporate significant forces, and by their
inappropriate use of collisionless fluid mechanics.  The ill-posed Jeans 1902
theory (neglecting non-acoustic density perturbations) is far too limited a basis
for discussion of structure formation because it neglects viscous forces,
turbulence forces, non-acoustic density fluctuations, particle collisions, and
the effects of diffusion on gravitational structure formation.  Jeans did linear
perturbation stability analysis (neglecting turbulence) of Euler's equations
(neglecting viscous forces) to reduce the problem of gravitational instability
of a nearly uniform ideal gas with density
$\rho$ only a function of pressure (the barotropic assumption) to one of
gravitational acoustics.  Furthermore, to reconcile his equations with the
linearized collisionless Boltzmann's equations and the resulting Poisson's
equation for the gravitational potential \citep{bt}, Jeans assumed the density
was zero in a maneuver appropriately known as the ``Jeans swindle''.    The
critical wave length for stability with all these questionable assumptions is the
Jeans length scale
$L_J$ given in Eq. 1 (see Table 1).

Density fluctuations in natural fluids are generally not barotropic  as
assumed by Jeans 1902 except in small regions and short times near powerful
sound sources, but are dominated by non-acoustic (non-barotropic) density
variations from turbulent mixing of temperature or chemical species
concentrations \citep{gib01}.  Even in the context of Jeans' theory (without
the Jeans swindle), any gravitational condensation on an acoustical density
maximum rapidly converts to a non-acoustical density maximum  because the
accreted mass retains the momentum of the presumed motionless ambient gas.
At this point one faces the more realistic problem of gravitational structure
formation in a moving, viscous, diffusive, possibly turbulent gas that was
addressed by Gibson 1996. Turbulence or viscous forces will dominate
gravitational forces at small distances from a non-acoustic point of maximum or
minimum density, but gravitational forces will dominate turbulent or viscous
forces at larger distances if the gas does not diffuse away faster than it can
condense or rarify due to gravity.

The Jeans 1902 analysis fails because the problem was ill posed. 
Self-gravitational structure formation in nature occurs on non-acoustic density
maxima and minima, not sound wave crests, but Jeans started by assuming a uniform
density, and was forced to set the density to zero in the Jeans swindle to avoid
the nonlinear problems of non-acoustic density maxima forming at acoustic
wavecrests.  The geometry of evolving non-acoustic density fields at scales
$L
\le L_J$ is
shown in Figure 1, with (top) and without (bottom) self-gravitational 
instability.  Without self-gravity a quasi-equilibrium at the 
Batchelor
length scale (Table 1) exists, where the competition between diffusive smoothing
of density gradients and strain rate steepening produces a universal turbulent
mixing scale
$L_B
\equiv (D/\gamma)^{1/2}$ at points of maximum and minimum density \citep{gib68}
that begin to decay by diffusion as soon as turbulence scrambles a uniform
density gradient to form such extrema with zero density gradients, as shown by
experiments and numerical simulations
\citep{gar88}.  Without gravity, points of maximum density always decrease by
diffusion and points of minimum density increase, always moving and diffusing
back toward the original configuration with uniform gradient and no density
extrema.

Evidence of primordial turbulence in CMB anisotropies \citep{bs02}
suggests non-acoustic density maxima necessary to seed primordial gravitational
structures were available in the plasma epoch.  Assuming the same initial
density conditions as Fig. 1 (top), the evolution of
density with self gravity (bottom) is very different.  The reason is that gravity
has the effect of causing a negative diffusivity
$D_{\rho G}(L) = -L^2 (\rho G)^{1/2}$ for distances $L < (D^2 /\rho G)^{1/4}$
close to the points of zero density gradient and maximum or minimum density
value.  Negative density diffusivity causes absolute gravitational instability at
non-acoustic density extrema.  
With self gravity (bottom) density maxima increase their density values toward
infinity and density minima decrease their density values toward zero.  Both
experience exponential growth to form condensates and voids for all length scales
$L \le L_J$ unless limited by turbulent, viscous or diffusive effects 
at the appropriate
Schwarz length scales.

Consider a well posed problem in self-gravitation.  A mass perturbation
$ M(0)$ with scale $L \ll L_J$ suddenly placed in a motionless fluid of
constant density
$\rho$ at time
$t = 0$ will increase as $ M(t) =  M(0) exp[2 \pi \rho G t^2]$
unless viscous, turbulent, other forces or diffusion arise to prevent the
exponential growth \citep{gib00}.  Everything slowly begins to move toward or
away from the mass perturbation for $t > 0$  and everything happens at once at
$t \approx t_{FF} = (\rho G)^{-1/2}$.   For $t < t_{FF}$ the radial velocity
$v_r \approx - G  M(0) t / r^2$ from Newton's law of gravity, so the mass
flow $d  M(t)/d t
\approx -
\rho 4
\pi r^2 v_r
\approx 4
\pi
\rho G  M(t) t$  and density $\rho$ are independent of radius.  Thus,
$d
 M(t) /
 M(t) \approx 4 \pi \rho G t$, so $ M(t) =  M(0) exp(2
\pi \rho G t^2)$. Since everything happens within
$L$ of the perturbation at
$t
\approx t_{FF}$, any ``pressure support'' mechanism is confined to this small
space-time region and has no affect on the weakly accelerated fluid that
slowly gains momentum at scales
$L \le L_J  \le x
\le ct$ and $L \le x
\le L_J$ for times $t \le t_{FF}$.  Since it takes a
time $t
\approx t_{FF}$ for a sound wave to reach a distance $x \approx L_J$ in time
$t \approx t_{FF}$, any hypothetical acoustic ``pressure support'' mechanism
of the Jeans criterion would be too late to prevent the exponential density
singularity at
$t
\approx t_{FF}$.  Note that the time required for self-gravitational structure
formation depends almost entirely on the mean density
$\rho$ and only very slightly on the size of the density perturbation
$\delta \rho \approx  M(0) L^{-3}$.

The diffusion velocity is
$D/L$ for diffusivity
$D$ at distance $L$ and the gravitational velocity is $L (\rho G)^{1/2}$.
The two velocities are equal at the diffusive Schwarz length scale
\begin{equation} L_{SD} \equiv [D^2 / \rho G]^{1/4}.\end{equation}  Weakly
collisional particles such as the hypothetical cold-dark-matter (CDM)
material cannot possibly form  potential wells for baryonic matter collection
because such particles have large diffusivity and will disperse, consistent
with observations \citep{sa02}.  Diffusivity
$D
\approx V_p \times L_C$, where $V_p$ is the particle speed and $L_C \approx 1/n
\sigma$ is the collision distance for particles with number density $n = \rho /
m$, $\sigma$ is the collision cross section, and $m$ is the particle mass.  CDM
particles have $\sigma_{CDM} \approx 10^{-40}$ m$^{2}$ or less compared to
baryonic
$\sigma_{B} \approx 10^{-20}$ m$^{2}$ values. Weakly collisional particles thus
have large collision distances $L_C$ and therefore large diffusivities $D$ and
large diffusive Schwarz lengths $L_{SD}$.  Therefore, the non-baryonic dark
matter (possibly neutrinos) must be the last material to fragment by self
gravity and not the first as assumed by CDM cosmologies.  In HGT  galaxy
formation
\citep{gib96}, the NBDM diffuses to  form either an outer halo of the galaxy
if the galaxy is isolated, or diffuses  completely away from the baryonic
galaxy to become part of a galaxy cluster halo or galaxy supercluster halo at
Mpc scales $ 3 \times 10^{22}$ m or more.

The baryonic matter is subject to large viscous forces, especially in the hot
primordial plasma from photon viscosity and gas states from large $V_p$ values
existing when most gravitational structures first formed, where $\nu \approx 
V_p \times L_C $ if $L_C \le L_H$.  The viscous forces per unit  volume
$\rho
\nu
\gamma L^2$ dominate gravitational forces $\rho^2 G L^4$ at small scales,
where
$\nu$ is the kinematic viscosity and $\gamma$ is the rate of strain of the
fluid.  The forces match at the viscous Schwarz length
\begin{equation} L_{SV} \equiv (\nu \gamma /
\rho G)^{1/2},\end{equation}
   which is the smallest size for self gravitational condensation or void
formation in such a flow.  Turbulent forces may require even larger scales of
gravitational structures.  Turbulent forces $\rho \varepsilon^{2/3} L^{8/3}$
match gravitational forces at the turbulent Schwarz scale
\begin{equation}L_{ST} \equiv \varepsilon ^{1/2}/(\rho G)^{3/4},\end{equation}
   where $\varepsilon$ is the viscous dissipation rate of the turbulence. Thus
in strongly turbulent regions with large $\varepsilon$ values one expects
large stars to form because the turbulent Schwarz scales are larger  than any
other hydro-gravitational scale.  In the cold, turbulent, dense molecular
clouds of the Milky Way Galaxy disk where the Jeans mass is near $M_\sun$ or
less, few stars form compared to the rate expected by the Jeans criterion
because
$L_{ST} \gg L_J$.

The Jeans criterion is nearly irrelevant to gravitational structure
formation triggered by non-acoustic density maxima and minima where
structures can form at scales that are either larger or smaller than $L_J$.
By Jeans' criterion no structures can form in the primordial plasma because
$L_J$ during  this hot, dense, epoch is always larger than the scale of
causal connection $L_H
\equiv ct$, where
$c$ is the speed of light and $t$ is the time since the big bang.  However,
the viscous and turbulent Schwarz scales $L_{SV} \approx L_{ST}$ became smaller
than
$L_J$ during this period, and first matched the more rapidly increasing
$L_H$ at time $t \approx 10^{12}$ seconds \citep{gib96}, well before $10^{13}$
seconds which is the time of plasma to gas transition (300,000 years).
Because the expansion of the universe inhibited condensation but enhanced
void formation in the weakly turbulent plasma, the first structures were
proto-supercluster-voids. At $10^{12}$ s the Reynolds number was near
critical so that
\begin{equation} (L_{SD})_{NBDM} \gg L_{SV} \approx L_{ST} \approx 5 \times L_K
\approx L_H = 3 \times 10^{20} \rm m, \end{equation}  where $L_{SD}$ applies
to the non-baryonic component and $L_{SV}$, $L_{ST}$, and $L_{K}$ apply to the
baryonic component.

As proto-supercluster fragments formed the voids between filled with
non-baryonic matter by diffusion, inhibiting further structure formation by
decreasing the gravitational driving force.  The baryonic mass density
$\rho
\approx 2
\times  10^{-17}$ kg/$\rm m^3$ and rate of strain
$  \gamma \approx 10^{-12}$ $\rm s^{-1}$ were preserved as hydrodynamic
fossils within the proto-supercluster fragments and within proto-cluster and
proto-galaxy objects resulting from subsequent fragmentation as the photon
viscosity and
$L_{SV}$ decreased prior to the plasma-gas transition and photon decoupling
\citep{gib00}.  As shown in Eq. 5, the Kolmogorov scale $L_K \equiv [\nu^3
/\varepsilon ]^{1/4}$ at the time of first structure nearly matched the
horizon scale $L_H
\equiv ct$ and the viscous and turbulent Schwarz scales, freezing in the
density, strain-rate, and spin magnitudes and directions of the proto-cluster
and proto-galaxy fragments of proto-superclusters.  Remnants of the
strain-rate and spin magnitudes and directions of the weak turbulence at the
time of first structure formation are forms of fossil vorticity turbulence
\citep{gib99b}. Thus HGT naturally explains observed bright galaxy spin
alignments and close angular associations with each other and quasars
without assuming intrinsic red shifts and mutual ejections.

The quiet condition of the primordial gas is  revealed by measurements of
temperature fluctuations of the cosmic microwave background  radiation, which
show an average $\delta T/T \approx 10^{-5}$, much too small for any
turbulence to have existed at that time of plasma-gas transition. 
This is to be expected because any turbulent plasma motions would have been
strongly damped by buoyancy forces after the first gravitational fragmentation
time
$10^{12}$ s.  Viscous forces in the plasma are inadequate to explain the lack
of primordial turbulence (kinematic viscosity $\nu$
$ \ge 10^{30}$ m$^2$ s$^{-1}$ is required versus only $4 \times 10^{26}$ m$^2$
s$^{-1}$ from Gibson 2000).  The gas temperature, density, viscosity,
and rate of strain are all precisely known at transition, so the gas viscous
Schwarz  mass
$L_{SV}^3 \rho$ is
$\approx 10^{24}$ kg, the mass of a small planet, or
$\approx 10^{-6} M_{\sun}$.  From HGT,
soon after the cooling primordial plasma turned to gas at $10^{13}$ s (300,000
yr), the entire baryonic universe condensed to a fog of planetary-mass
primordial-fog-particles (PFPs).  These gas-cloud objects gradually cooled,
formed H-He rain, and eventually froze solid to become the baryonic dark
matter and the basic material of  construction for stars and everything else,
$\approx 30 \times 10^{6}$ rogue planets per star.  The PFP mass $\approx
10^{-6}M_{\sun}$ is above the evaporation stability limit
$\approx
10^{-7}M_{\sun}$ \citep{der92}.

The Jeans mass $L_J^3 \rho$ of the primordial gas at transition was
$\approx 10^6 M_{\sun}$, the mass of a
globular star cluster.  Proto-galaxies fragmented at the PFP scale but also
at this proto-globular-star-cluster PGC scale
$L_J$, although not for the reason given by the Jeans 1902 theory.  Density
fluctuations in the gaseous proto-galaxies were absolutely unstable to  void
formation at all scales larger than the viscous Schwarz scale $L_{SV}$.
Pressure can  only remain in equilibrium with density without temperature
changes in a gravitationally expanding void on scales smaller than the Jeans
scale.  From the second law of thermodynamics, rarefaction wave speeds that
develop as density  minima expand due to gravity to form voids are limited to
speeds less than the sonic velocity.  Cooling would therefore occur and be
compensated by radiation in the otherwise isothermal primordial gas when the
expanding voids approached the Jeans scale. Gravitational fragmentation of
proto-galaxies will then be accelerated by radiative heat transfer to these
cooler regions, resulting in fragmentation at the Jeans scale and isolation
of proto-globular-star-clusters (PGCs) with the primordial-gas-Jeans-mass.

These
$10^{36}$ kg PGC objects were not able to collapse from their own self
gravity  because of their internal fragmentation at the viscous Schwarz scale
to form $10^{24}$ kg PFPs. The fact that globular star clusters have
precisely the same density and primordial-gas-Jeans-mass from galaxy to
galaxy proves they were all formed simultaneously soon after the time of the
plasma to gas transition $10^{13}$ s.  The gas has never been so uniform
since, and no mechanism exists to recover such a high density, let alone such
a high uniform density, as the fossil turbulent density value $\rho \approx 2
\times 10^{-17}$ kg/$\rm m^3$.  Young globular cluster formation in BDM halos
in the Tadpole, Mice, and Antennae galaxy mergers show that dark PGC clusters
of PFPs are remarkably stable structures, persisting without disruption or
star formation for more than ten billion years.

So the proto-galaxies (PGs) before stars were rapidly fragmented at both the
Jeans and viscous Schwarz scales to form PGCs, a million per galaxy, with
embedded PFPs, a trillion per PGC and $10^{18}$ per PG.  Viscous forces
inhibited the formation of larger planetoids by accretion of the PFPs except
at the cores of the PGCs and the cores of the PGs.  It was a very quiet and
gentle time.  The gentleness of this time of first star formation is reflected
in the small, uniformly distributed, long-lived, Population II stars of the
ancient globular-star-clusters observed in the Milky Way Galaxy.  The time to
form the first of these stars was probably one million years (the primordial
density free fall time), not the 275 million ``dark age'' years required by
the standard CDM model while the CDM halos formed by frictionless clustering
of CDM seeds.  Open star clusters formed from captured and disrupted PGCs in
the Galaxy disk are exposed to higher levels of radiation and tidal friction
than PGCs in the BDM halo, and form large irregularly distributed stars. The
large sizes of the disk stars and their slow rate of formation reflect the
high turbulence levels of disk star forming regions (from supernovas)
compared to the low turbulence levels existing where stars are formed in
isolated PGCs.  From HGT, the thin disks of spiral galaxies are accretion
disks forming within large, massive, spherical BDM halos surrounding the
original smaller denser
$\approx 10^{20}$ m galaxy cores from which the PGCs diffused, rather than
collapsed gaseous pancakes in CDM halos as envisaged in CDM standard models.
Relatively low gas levels are observed in spiral galaxy cores according to
HGT, where excess gas is frozen out on the ambient PFPs.

With the expansion of the universe and further cooling, the PGCs and PFPs
cooled as well, reaching the 14 K freezing point of hydrogen at about a
redshift of 30 at a time about 80 My.  Because most of the gas is isolated in
individual PFPs separated from their neighbors by distances of about
$10^{14}$ m corresponding to the primordial density, and because most PFPs
within a PGC are initially surrounded homogeneously by identical objects with
the same spacing,  the tendency without external forcing would be for them to
slow down due to  friction from the inter-PFP gas and remain isolated
indefinitely.  After freezing, the  PFP collision cross section will decrease
but the inter-PFP gas density and friction will also decrease, permitting
speedup of the PFPs and a new meta-stable state of equilibrium.  The
Gunn-Peterson trough phenomenon shows intergalactic gas effectively vanishes
for red shifts smaller than 6 ($t \approx$ 700 My) in quasar spectra  with
their Lyman-$\alpha$ forests of intervening galaxies.  This suggests that
most baryonic-dark-matter galaxy halo gas has frozen out on  PFPs embedded in
the BDM halos at times less than a billion years after the big bang.

As the inter-PGC gas density decreases with time since PFP creation, these
PGC clumps of condensing baryonic dark matter become increasingly
collisionless and diffusional.  Those formed near the core of the
proto-galaxy have a higher likelihood of mutual capture, interaction,
disruption, and conversion to stars, but those formed on the periphery of the
proto-galaxy will tend to either capture each other to form
proto-dwarf-galaxies or diffuse to larger and larger orbits.  The inner PGCs
should be the most agitated by tidal forces and therefore the most likely to
form stars and eventually the full population of
$10^{6}$ stars observed in a globular star cluster.  Those further out might
form only a few thousand stars and drift to the outskirts of the luminous
portion of the galaxy,  sometimes being captured or disrupted by the disk,
but the largest number diffusing farther out to form the baryonic-dark-matter
(BDM) halo of the galaxy, invisible  unless  disturbed by intruders as in the
case of the Tadpole merging galaxy system (VV29=Arp 188=UGC 10214).  Figure 3
shows the HST/ACS image with its combination of wide field and high
resolution, which clearly reveals the existence of star wakes, a dark dwarf
galaxy and young-globular-clusters triggered into existence from the dark
baryonic-dark-matter halo by the merging cluster of galaxy components VV29cdef
as they spiral in to the central galaxy VV29a and merge \citep{gs03b}.

Sparse globular clusters such as Palomar 5 with luminous mass $10^{4} M_\sun$
and Palomar 13 with luminous mass $10^{3} M_\sun$ observed in the Milky Way are
likely dim rather than disrupted PGCs with 98\% to 99.9\% of their mass intact as
dark-matter PFPs.  Nearly half of the stars of Pal 5 are observed as long
tidal tails along its orbit about the Galaxy center
\citep{roc02}, suggesting that the same tidal forces that produced the tails
may also be producing the stars from a large supply of remaining PFPs.  The
amount of dark matter in dim globular clusters like Pal 5 and Pal 13 in the
Milky Way, and in Galactic dwarf spherical galaxies (dSphs), is likely
underestimated in virial estimates $M
\approx (Rv^2 /G)$ of the dark mass $M$
  because friction from PFP gas is neglected that reduces the velocity
variance $v^2$ of the stars within radius
$R$. A mass-to-light ratio  $\Upsilon \approx 40$ for Palomar 13 was recently
reported
\citep{Cot02},  which is the first estimate from
$v^2$ measurements in dim globular star clusters and is much higher than
$\Upsilon
\approx 2$ expected.  In another estimate (Cote et al. 2002 Fig. 14b)
$\Upsilon \approx 7000$ fitting to the luminosity relation for dSphs which is
closer to the value $\Upsilon \approx 1000$ expected from HGT.   Large
$\Upsilon$ values measured in dSph galaxies have been assumed to reflect
non-baryonic CDM halos
\citep{Cot02}, but from Tadpole we see PGC and dSph dark matter is mostly
baryonic because it forms stars when agitated.

The hydro-gravitation-theory (HGT) scenario presented here is significantly
different from standard CDMHC cosmology based on Jeans 1902 
precepts. Gravitational structures in the baryonic matter fragmented during the
plasma epoch with supercluster to galaxy masses.  The protosuperclusters  have
the observed uniform supercluster mass of about $10^{47}$ kg (including their
non-baryonic halos), with a uniformity not expected for hierarchical clustering
of CDM halos assumed in the standard CDMHC model.  Weak primordial turbulence is
damped by buoyancy forces of the first structures to form fossil vorticity
turbulence
\citep{gib99b}.  The density and rate-of-strain values at the time of first
structure ($10^{12}$ s) are preserved (fossilized), and can still be detected in
the density of globular star clusters and the masses of PFPs.  The non-baryonic
matter diffused to fill the protosuperclustervoids, limiting the amplitude of
the density contrast formed to the small values observed in the CMB. 
Superclusters, clusters and galaxies never contracted or collapsed but continue
to fragment and gently expand at slower rates than the rest
of the universe.

\section{Evidence Supporting the Hydro-Gravitational Theory}

Evidence supporting the hydro-gravitational theory of structure formation is
gradually accumulating.  High resolution telescope images in star  forming
regions and supernova remnants are full of poorly explained small objects,
knots and ``ejecta'' that are likely to be evaporating PFPs. Cometary
globules  and the huge masses of gas observed by the HST near hot dying stars
in planetary nebulae are much more easily explained as the result of
evaporating ambient PFPs in the interstellar medium than they are as ejecta.
Plasma beams from young stars produce Herbig-Haro objects with cometary
tails, and the plasma beams  themselves pulsate as though the accreted
material forming the star contained  planetary mass chunks.  Numerous brown
dwarfs and Jupiters have been detected whose origin is mysterious without
HGT.  The total mass of extra-solar planets discovered increases with
decreasing planet-mass, even though low mass extra-solar planets are more
difficult to detect, as expected if extra-solar planets are formed by
accretion of ambient  PFPs before their capture by stars.

The closest planetary nebula is Helix, Figure 2, where some 6500  cometary
globules have been identified, each with mass about
$10^{25}$ kg and size $10^{13}$ m, and separated by about $10^{14}$ m
\citep{ode96}. Radial tails point
away from the hot, central, white dwarf star at $\approx 4
\times 10^{15}$ m. The mass $\approx 2 \times 10^{25}$ kg and  separation
distances
$\approx 10^{14}$ m of the Helix cometary globules give a halo density $\rho
\approx 10^{-17}$ kg m$^{-3}$, suggesting that PFPs originally in a sphere
with radius
$\approx 4
\times 10^{15}$ m have accreted to form the central star out of an ISM with
mass dominated by PFPs, with primordial density
$\rho
\approx 10^{-17}$ kg m$^{-3}$, leaving PFPs at the hole boundary 
(Oort Cloud) evaporated
and illuminated by the hot dying star.  Measured evaporation rates \citep{mea98}
are too high for the globules shown in Figure 2 to be PFPs, so they must be JPPs
with at least Jupiter mass to have survived this long \citep{gs03c}.  If the ISM
surrounding the Helix represents a dark proto-globular-star-cluster (PGC), the
PN represents the luminous boundary of a hole formed by accreted PFPs to form
the illuminating central star, with mass
$\approx 10^{-5}$ of the surrounding ISM PGC baryonic dark matter cloud of
PFPs.  The gas in the nebula is of order the mass of the star, but
the total mass of the atmosphere of the dying star before it is ejected
should be less than about
$10^{-7} M_{\sun}$.   

The mass  density for the Helix globules
$10^{-14}$ kg/$\rm m^3$ is three orders of magnitude larger than the density
of any atmosphere ejected by the white dwarf star.  Shock waves can increase
densities by no more than factors $\approx 6$, not the factors $\approx
10^{7}$ required to explain the objects of Figure 2.  These cometary globules
are usually explained as due to Rayleigh-Taylor instability, where a  dense
gas shell ejected first is accelerated by a lighter shell ejected more rapidly
later, but RT instability cannot increase the density of the dense
shell by many orders of magnitude or decrease the density in the RT fingers
to  form wakes as observed.  RT fingers only form in a narrow range of
Reynolds numbers near critical, but this is not the case for the Helix
objects.  If the cometary globule separations in Helix reflects that of the
ambient PFP population, the mass density of PFPs is near the value
$10^{-17}$ kg/$\rm m^3$ expected for a PGC.  The PGC may have supplied the
PFPs from which the Helix star was formed, consistent with the size of the $4
\times 10^{15}$ m region surrounding the white dwarf that is observed to be
empty of cometary globules \citep{gs03c}.

However, the most conclusive and convincing evidence for the PGC-PFP
HGT scenario is still the technically challenging one of
quasar microlensing.  The Schild 1996 interpretation that the QSO 0957+561A,B
quasar lens galaxy must be dominated by planetary mass objects is based on the
observed high twinkling frequency of the two quasar images A,B.  The initially
controversial Schild time delay has now been confirmed by several independent
observers, and is in the process of further refinement by an international
team of observers using telescopes at sufficient points on the Earth to
provide continuous coverage and independent confirmation of the twinkling QSO
0957+561A,B light curves (Colley et al. 2002, 2003).   At this time four other
lensed quasars have been observed with sufficiently precise time delays of
the mirage images to extract the planetary mass twinkling frequencies, and
these have been observed and reported (Burud et al. 2000, 2002; Hjorthe et
al. 2002, Schechter et al. 2002). The specific microlensing profile
discovered in Q0957 by Colley and Schild (2003) with high statistical
significance shows that the lens galaxy G1 halo population has a
significant optical depth of planetary mass objects, and the double-ring
halo model of Schild and Vakulik (2003) shows how such rapid microlensing
plausibly follows from simulations.

Evidence of substructure in several lensing galaxies is reported by Dalal and
Kochanek 2002 from the difference in brightness of the background  source
mirages, with substructure masses in the range
$10^6
\-- 10^9 M_{\sun}$.  This evidence is presented to suggest that the missing
CDM halos of the local group of galaxies is not a crisis for the hierarchical
CDM model of galaxy formation that overestimates the numbers of dwarf
galaxies by orders of magnitude compared to observations because the actual
galaxy substructures may be dark.  However, the substructure evidence
\citep{dal02} of dark non-baryonic CDM halos may also be interpreted as
evidence that significant fractions of the lens galaxies central masses are
in dark baryonic  clumps.   Dark PGC clumps of PFPs dominating central galaxy
masses is predicted by HGT.
Clumps of dark PGCs triggered to form clumps of YGCs by the star formation
wake of VV29c are indicated in the VV29b filament of Tadpole by the HST/ACS
images
\citep{tra03}, Figure 3.

Merging galaxy systems such as the Tadpole, Mice, and Antennae are
characterized by the appearance  out of the dark of young-globular-clusters
(YGCs) and large quantities of gas.  Using the HST and a wide variety of
other telescopes to cover a  broad range of frequencies
\citep{zha01}, nearly a thousand YGCs have been  identified and studied in
the Antennae system (NGC 4038/9), which is the closest  merging galaxy system
at only
$6
\times 10^{23}$ m.  At this distance the clusters are well resolved.   The
mass function inferred from the luminosity for the young (5 Myr) globular
clusters in Antennae
\citep{fall01} is very different than that for old ($\ge 10$ Gyr) globular
clusters (OGCs), falling with power law slope -2 from a maximum near $10^{4}
M_\sun$ rather than showing the narrow lognormal peak near $2 \times 10^{5}
M_\sun$ widely observed for OGCs.  The brightest YGCs have the same mass as
the OGCs.  We suggest that the low masses of the most numerous YGCs is simply
because most of the mass of the PGCs from which these YGCs are forming is
still tied up in dark matter PFPs that are rapidly accreting to form stars
under the influence of the tidal merger.

Figure 4 shows an HST image of Stephan's Quintet  (SQ, HGC 92, Arp 319, VV
288) which is number 92 in the Hickson catalog of compact clusters. Like over
40$\%$ of the HGCs, the SQ galaxies \citep{bur61} have highly discordant
redshifts.  These are easily explained by HGT, where the SQ and other HGC
galaxies with discordant redshifts represent galaxy clusters that have gently
separated only recently due to the expansion of the universe \citep{gs03a}.  The
star wakes in Figure 4 show luminous evidence of the baryonic dark matter
cluster halo and its boundary.  No evidence of a galaxy merger is shown, in
contrast to the Tadpole merger in Figure 3 where the merged galaxy is revealed. 
Numerous young globular clusters (YGCs) have been identified in SQ \citep{gal01},
confirming the HGT prediction that the baryonic dark matter consists of PGC
clumps of PFPs that can be triggered to form
stars by radiation and tidal forces.

Figure 5 shows the tomographic mass distribution of the rich galaxy cluster Abell
1689 determined from the gravitational distortion of 6000 background galaxy
images by 4000 foreground galaxies in the cluster \citep{tys95}.  A mass to
light ratio of about 400 suggests the cluster halo is non-baryonic dark matter
with an extent determined by the NBDM particle diffusivity $D_{NBDM}$ and the
large cluster density.  The Schwarz diffusive scale $L_{SD} \equiv (D_{NBDM}^2
/\rho G)^{1/4}$ from the mass profile is
$\approx 10^{22}$ m, giving
$D_{NBDM}
\approx 6 \times 10^{28}$ m$^2$ s$^{-1}$ from the cluster density $5 \times
10^{-21}$ kg m$^{-3}$.  Such a large diffusivity excludes any possibility that
the outer dark matter halo is baryonic.  It fixes the value of the $m/\sigma $
ratio to a value $\rho D (r/GM)^{1/2} \approx 10^3$ kg m$^{-2}$.  Because the
maximum collision cross section $\sigma$ for non-baryonic matter is about
$\le 10^{-36}$ m$^2$, the maximum particle mass for the NBDM is
$\le 10^{-33}$ kg, excluding WIMPs with $m \gg 10^{-27}$ kg as NBDM candidates by
large factors.  For example, the neutralino mass is about $10^{-24}$ kg, giving
a $\sigma$ value of $10^{-27}$ m$^2$, which is so large that neutralino-NBDM
would be easily detected.  The NBDM becomes  unstable to gravitational
fragmentation when
$L_{SD}$ becomes less than the horizon scale $L_H$.  This occurs at $t \approx
(D^2 /\rho G)^{1/4} c^{-1}$, soon after the plasma-gas transition at 
300,000 years, for the indicated diffusivity.  Outer halos scale with $M^{-1/4}$
so the galaxy NBDM outer-halo-radius is $\approx 1.5 \times
10^{22}$ m, or about four times the observed BDM inner-halo-radius for the
Tadpole galaxy
\citep{gs03b}. Since baryons can be no more than $\approx 1/30$ of the mass of
a flat universe from nucleosynthesis, non-baryons can be no more than $\approx
30$ times the mass of baryons.  Thus (neglecting any $\Lambda$ contribution)
galaxy outer halo densities are dominated by non-baryons and galaxy inner halo
densities are dominated by baryons.

\section{Conclusions}

High resolution images of the Helix Nebula, the Tadpole merger and the Stephan
Quintet (SQ) compact cluster support the hydro-gravitational theory (HGT)
prediction that primordial-fog-particles (PFPs) in
proto-globular-star-clusters (PGCs) dominate baryonic dark matter (BDM)
inner-galaxy-halos and the interstellar medium (ISM).  Young globular clusters
and star wake alignments in Tadpole and SQ are explained by HGT and are not
explained by previous models.  
Evidence accumulates showing high frequency twinkling of quasar images lensed
by foreground galaxies, confirming that the galaxy missing baryonic  mass is
dominated by PFP mass ``rogue planets'' \citep{sch96}.

Large numbers of extra-solar planets have been discovered with Jupiter mass
which must be made of H-He gas, but which are shown to be impossible by the
Jeans 1902 gravitational structure formation criterion in $\S1$.  Such planets
cannot be formed by condensation on rocky cores because the formation of
rocks is also impossible by the Jeans criterion.  Jupiters and rocks form
naturally using hydro-gravitational theory \citep{gib96} by accretion of PFPs
to form larger planets and stars, and by gravitational accumulation of the
dust of supernovas by PFPs and larger planets to form rocky cores.  The Helix
Nebula and other planetary nebula show evidence that the ISM mass is dominated
by PFPs as predicted by HGT.  These accrete to form JPPs; Jupiters, brown dwarfs
and stars.  Evidence that the total mass of small extra-solar JPPs exceeds the
total mass of larger extra-solar JPPs, as expected from HGT, suggests that the
total mass of brown dwarfs in a galaxy exceeds  the mass of stars.
This prediction of HGT can be tested by the Space InfraRed Telescope Facility
(SIRTF) when it is launched, possibly this year.

Stephan's Quintet is an example of many compact clusters of galaxies with
discordant red shifts that refute CDMHCC and support HGT.  HST images confirm
the PGC-PFP baryonic dark matter halo model indicated by the Tadpole merger,
but show star wakes of cluster galaxies separating through the BDM cluster
halo rather than merging as shown by Tadpole and as predicted by HGT.  Angular
clustering of QSOs and bright galaxies \citep{bur03} also rule out CDMHCC and
support either HGT without new physics or intrinsic red shifts and QSO
ejection by AGN galaxies
\citep{arp01} and a possible failure of the big bang hypothesis \citep{hoy00}.
The forty year authoritative accumulation of inexplicable angular
clustering of galaxies, clusters and QSOs with anomalous red shifts 
are readily explained by HGT as the result of perspective and gentle straining
along lines of sight by the expansion of the universe, with small
transverse velocities reflecting the viscous origins and frictional effects of
fragmenting clusters in gassy, baryonic, galaxy and cluster halos.  

From HGT and the mass distribution \citep{tys95} computed for a
rich galaxy cluster we find the NBDM particle mass is too small to be a CDM
candidate, with fragmentation scales for non-baryonic-dominated
outer-galaxy-halos larger than radius values observed for baryon-dominated
inner-galaxy-halos.  Cusps at galaxy cores expected from small, clustered,
collisionless, CDM-halos are not observed
\citep{sa02}.    Hierarchical clustering cosmologies are contradicted by the
incidence of strong-lensing clusters
\citep{gla03}.  The most distant observable galaxies  are
already clustered \citep{sti00} and the closest clusters \citep{gs03c} are
fragmenting.  HGT and the preponderance of accumulated  evidence suggest that
cold-dark-matter, hierarchical-clustering-cosmologies (CDMHCCs), and the Jeans
1902 gravitational structure formation criterion are physically incorrect and
contrary to observation, and should be abandoned.

\acknowledgments The authors are grateful for many constructive suggestions
by the Editor and Referee.

\clearpage

\begin{figure}
         \epsscale{.6}
         \plotone{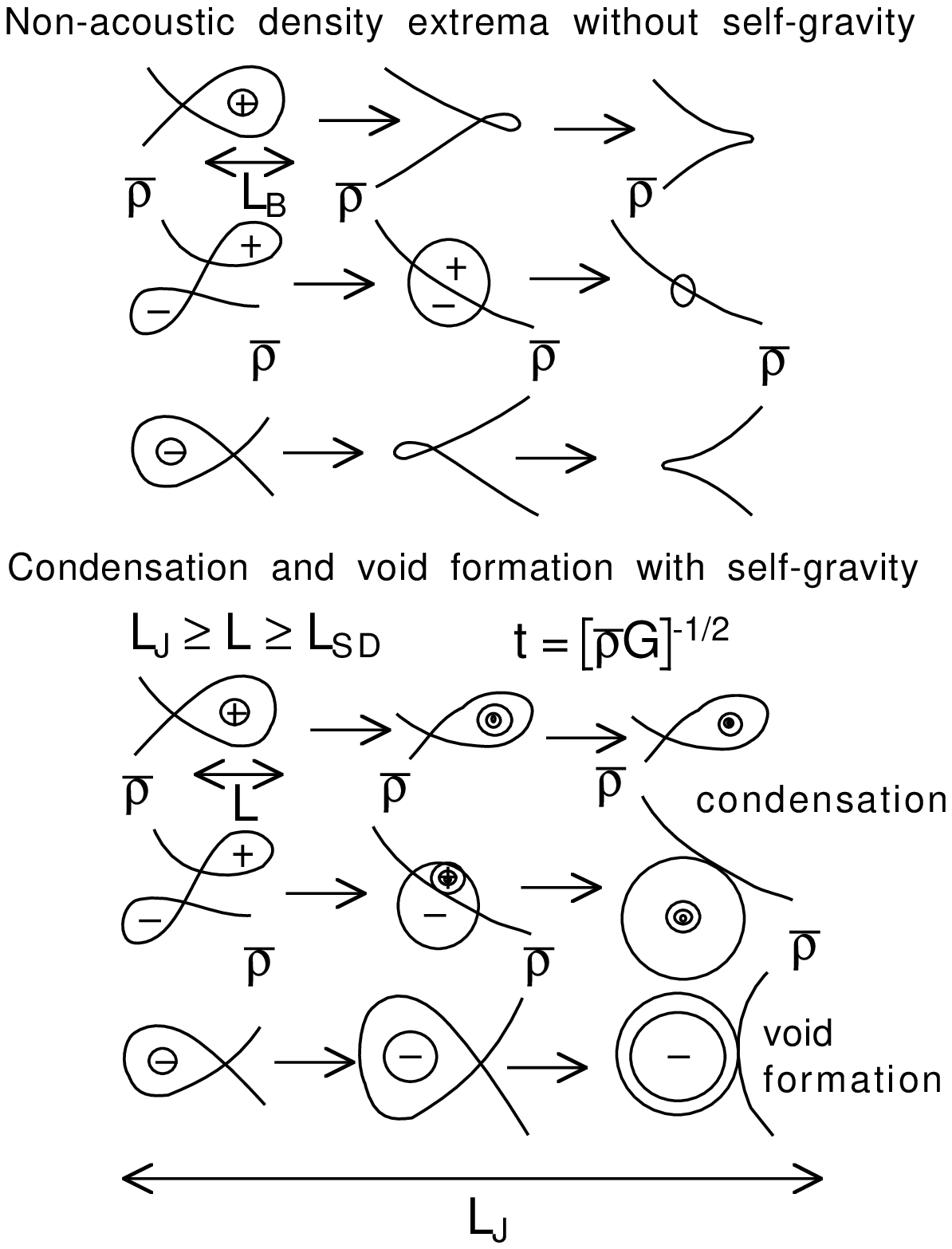}
         \caption{Schematic diagrams of non-acoustic iso-density surfaces
without self-gravitation (top) and with self-gravitation (bottom).  The mean
density is $\bar{\rho}$. Without self-gravitation turbulence
scrambles density to produce maximum points ($+$), minimum points ($-$) and
doublets ($+-$)  that diffuse to symmetric geometries and then
move with the fluid  \citep{gib68} at the Batchelor diffusive scale
$L_B$ (see Table 1).  With self-gravitation (bottom) on scales smaller than
the Jeans scale $L_J$ the effective diffusivity becomes negative for scales $L$
smaller than the diffusive Schwarz scale
$L_{SD}$. Density maxima and minima are absolutely unstable and
grow exponentially to form condensates or voids unless limited by turbulence,
viscosity or diffusion at the appropriate Schwarz scale $L_{ST}$,  $L_{SV}$,
or
$L_{SD}$, whichever is largest.  Note that for primordial non-baryonic cold
dark matter (CDM) $L_{SD} \gg L_J > L_H$; consequently no self-gravitational
(CDM-halo) structures can form.}
         \end{figure}

\begin{figure}
         \epsscale{0.8}
         \plotone{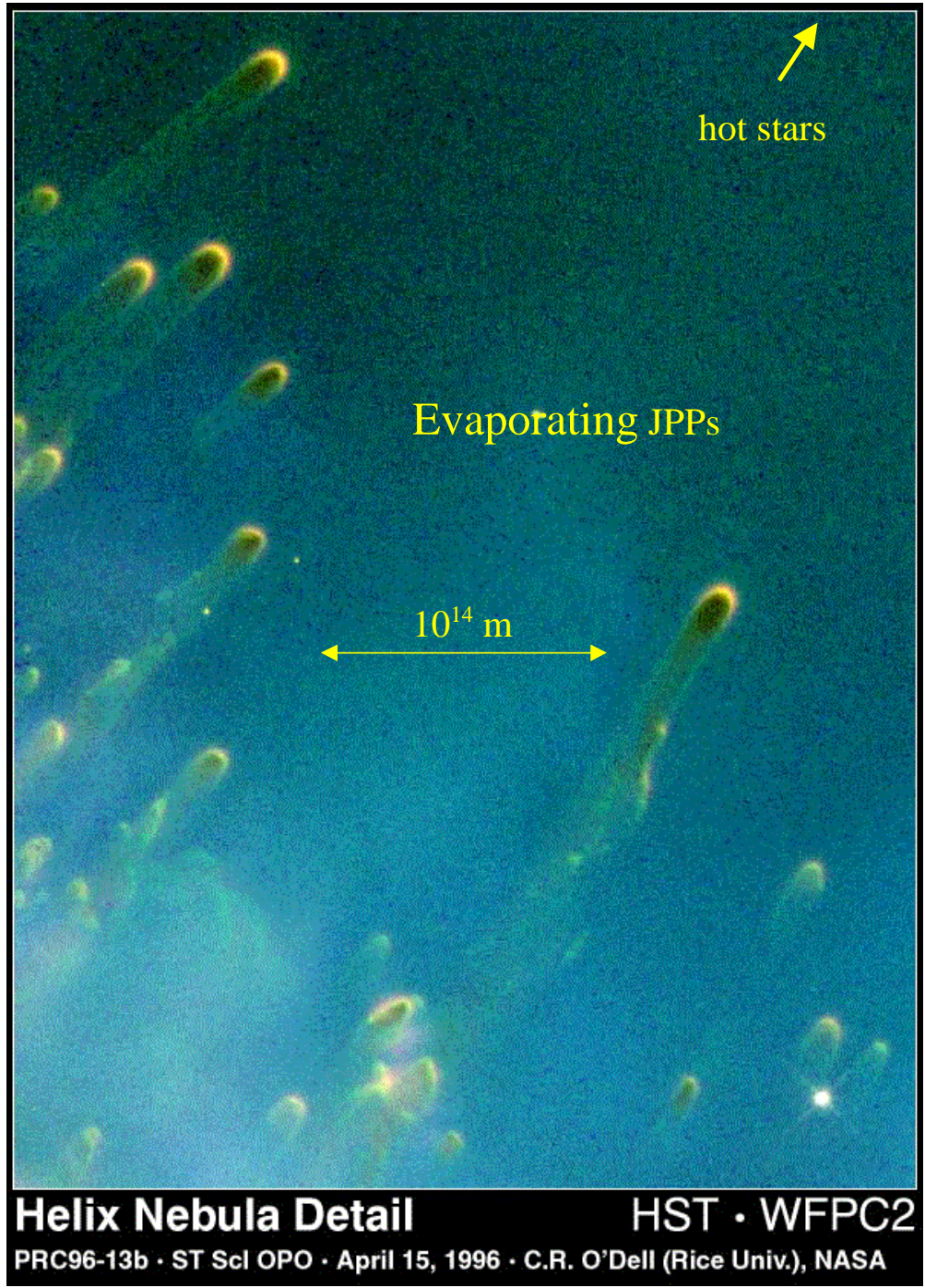}
         \caption{Evaporating JPPs observed by HST in the Helix Planetary
Nebula,
$4.5
\times 10^{18}$ m from Earth
\citep{ode96}. }
         \end{figure}

\begin{figure}
         \epsscale{1.0}
         \plotone{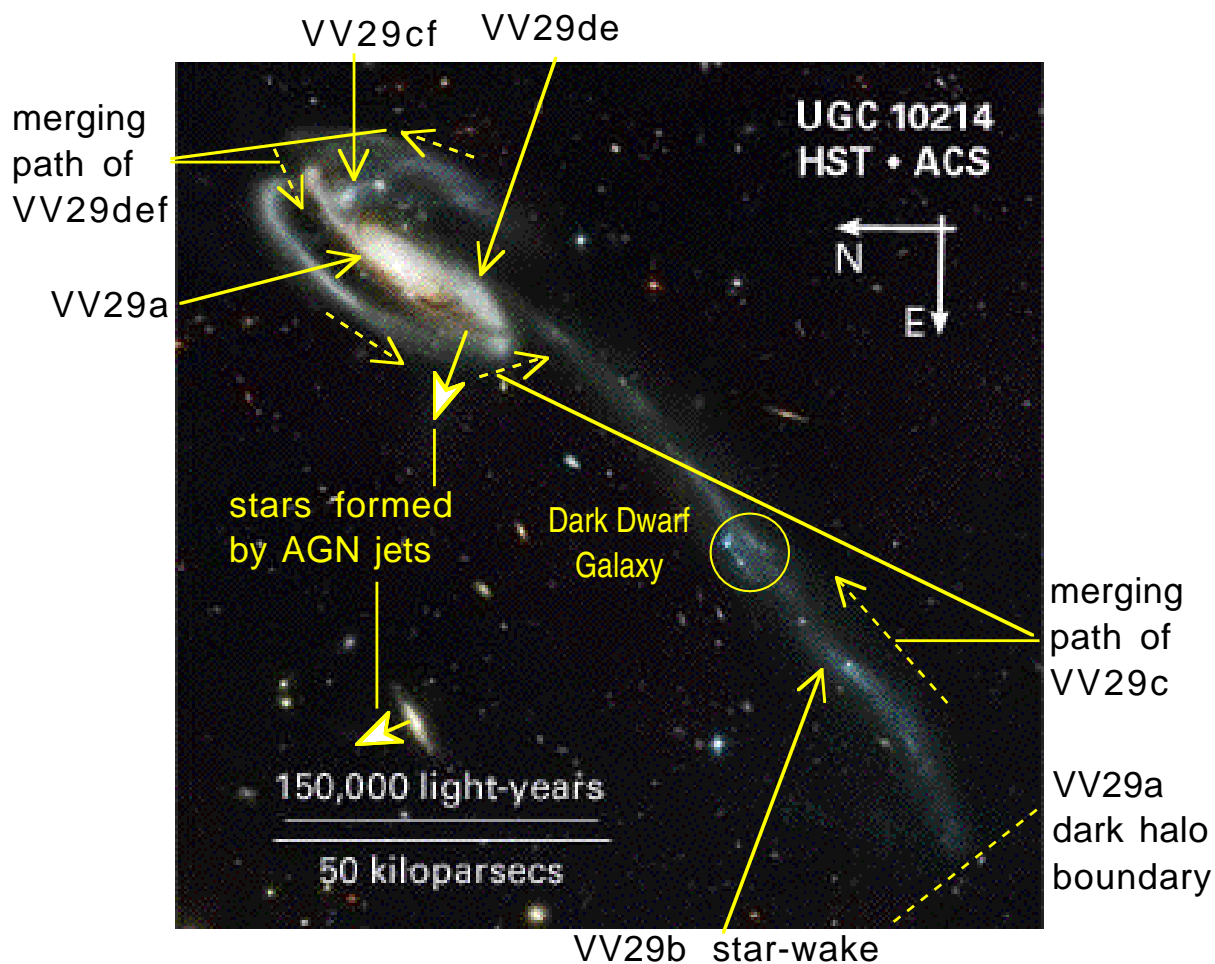}
         \caption{Interpretation of HST/ACS Tadpole image (April 30, 2002
press release) using the Gibson 1996-2000 hydro-gravitational structure
formation  cosmology, where the merging galaxy structures VV29cdef enter the
VV29a dark halo at lower right (dashed line) and merge along luminous trails
of star formation triggered by radiation and tidal forces acting on the PGCs
and PFPs of the dark halo of VV29a.  The VV29b galaxy is interpreted as a
star-wake of VV29cdef  merging galaxy components in the VV29a dark baryonic
halo, and not as any sort of  collisionless tidal tail.  Star formation
regions (visible in higher resolution images) triggered in the
baryonic-dark-matter by AGN jets of VV29e and a background galaxy are shown
by arrows.  A dark dwarf galaxy is revealed by a straight row of $\ge 42$ closely
spaced YGCs \citep{tra03} pointing precisely ($\le 1 \degr$) toward the VV29c
entry star-wake. North is to the left and East is to the bottom.}
         \end{figure}

\begin{figure}
         \epsscale{0.6}
         \plotone{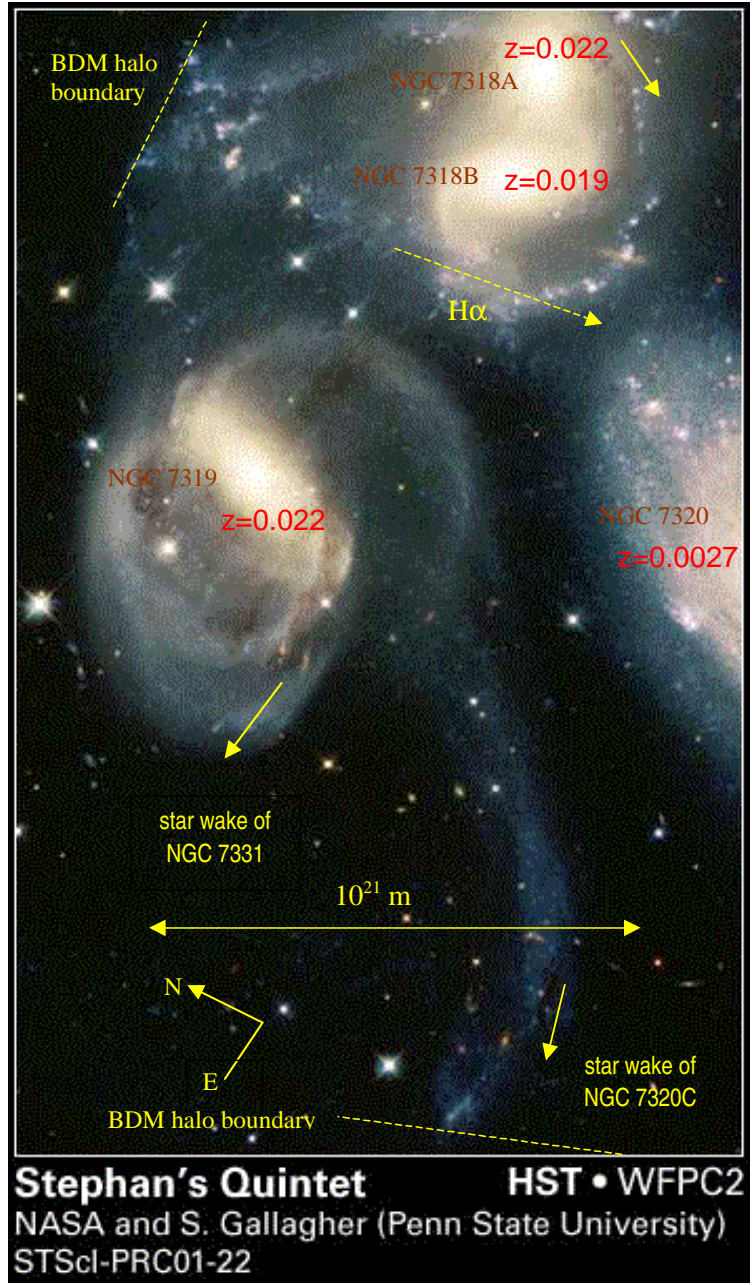}
         \caption{Hubble Space Telescope image of Stephan's Quintet. Dust and
star wakes (arrows) are produced as SQ related galaxies gently separate from
each other through the cluster baryonic-dark-matter (BDM) halo of PGCs and
PFPs, triggering star formation.  Star wakes of mergers and collisions are not
observed.  Widely differing red shifts ($z=0.022, 0.022, 0.019, 0.0027$) show
the galaxies are in a thin (1/700) pencil stretched by the expansion of the
universe \citep{gs03a}.  The galaxies fragmented recently and stay in the
line-of-sight pencil due to perspective and their small transverse velocities and
viscous-gravitational origin according to HGT, contradicting CDMHCC.}
        \end{figure}

\begin{figure}
         \epsscale{0.6}
         \plotone{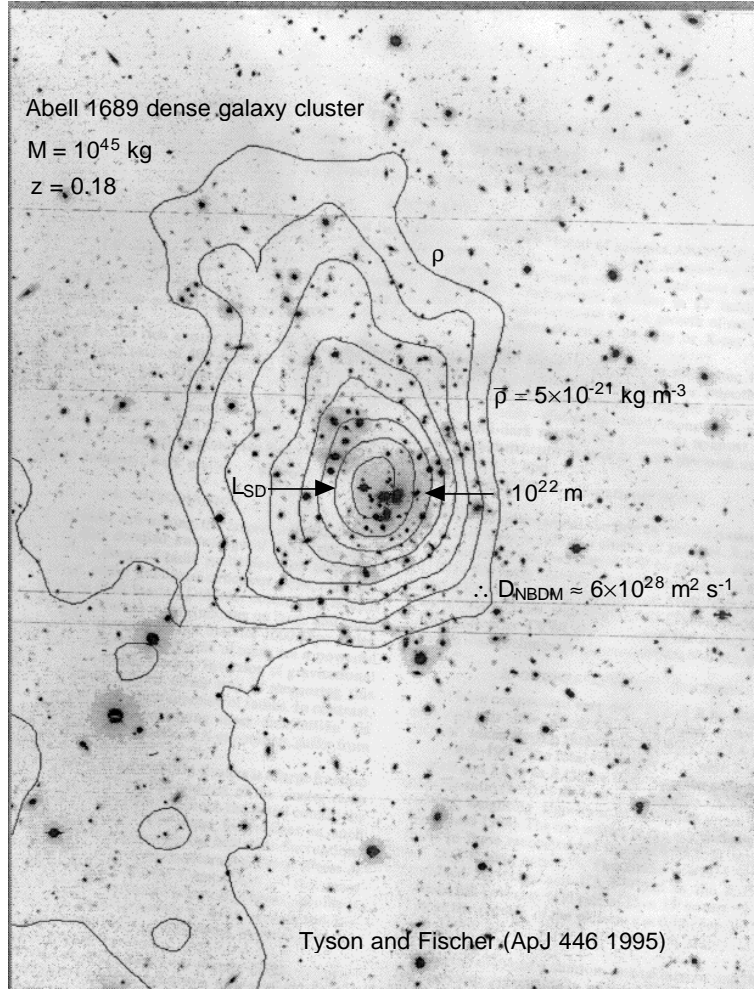}
         \caption{Mass distribution of rich galaxy cluster Abell 1689
determined by a tomographic analysis using gravitational distortions of 6000
background galaxies by 4000 foreground galaxies \citep{tys95}.  The mass of
the cluster is about
$10^{45}$ kg with redshift
$z = 0.18$ and average core density $5 \times 10^{-21}$ kg m$^{-3}$.  The
large mass to light ratio of 400 suggests that the dark matter halo is
non-baryonic.  Interpreting the halo size
$10^{22}$ m as the Schwarz diffusive scale $L_{SD}$ gives a large diffusivity
$D_{NBDM}
\approx 6 \times 10^{28}$ m$^2$ s$^{-1}$ that also suggests the halo is
non-baryonic.}
        \end{figure}

\clearpage

\begin{deluxetable}{lrrrrcrrrrr}
\tablewidth{0pt}
\tablecaption{Length scales of self-gravitational structure formation}
\tablehead{
\colhead{Length scale name}& \colhead{Symbol}           &
\colhead{Definition$^a$}      &
\colhead{Physical significance$^b$}           }
\startdata Jeans Acoustic & $L_J$ &$V_S /[\rho G]^{1/2}$& ideal gas pressure
equilibration

\\Jeans Hydrostatic & $L_{JHS}$ &$[p/\rho^2 G]^{1/2}$& hydrostatic pressure
equilibration

\\ Schwarz Diffusive & $L_{SD}$&$[D^2 /\rho G]^{1/4}$& $V_D$ balances $V_{G}$
\\  Schwarz Viscous & $L_{SV}$&$[\gamma \nu /\rho G]^{1/2}$& viscous force
balances gravitational force
  \\ Schwarz Turbulent & $L_{ST}$&$\varepsilon ^{1/2}/ [\rho G]^{3/4}$&
turbulence force  balances gravitational force
\\

Kolmogorov Viscous & $L_{K}$&$ [\nu ^3/ \varepsilon]^{1/4}$& turbulence
force  balances viscous force
\\

Batchelor Diffusive & $L_{B}$&$ [D / \gamma]^{1/2}$&
diffusion balances strain rate
\\

Collision & $L_{C}$&$ m \sigma ^{-1} \rho ^{-1}$& distance between particle
collisions
\\

Horizon, Hubble & $L_{H}$&$ ct$& maximum scale of causal connection
\\


\enddata
\tablenotetext{a}{$V_S$ is sound speed, $\rho$ is density, $G$ is Newton's
constant, $D$ is the diffusivity, $V_D \equiv D/L$ is the diffusive velocity
at scale $L$, $V_G \equiv L[\rho G]^{1/2}$ is the gravitational velocity,
$\gamma$ is the strain rate,
$\nu$ is the kinematic viscosity,
$\varepsilon$ is the viscous dissipation rate, $m$ is the particle mass,
$\sigma$ is the collision cross section, light speed $c$, age of universe
$t$.}

\tablenotetext{b}{Magnetic and other forces (besides viscous and turbulence)
are negligible for the epoch of primordial self-gravitational structure
formation considered here
\citep{gib96}.}


\end{deluxetable}

\clearpage

\begin{deluxetable}{lrrrrcrrrrr}
\tablewidth{0pt}
\tablecaption{Acronyms}
\tablehead{
\colhead{Acronym}& \colhead{Meaning}           &

\colhead{Physical significance}           }
\startdata

BDM & Baryonic Dark Matter&PGC clumps of PFPs from HGT
\\

CDM & Cold Dark Matter& questioned concept
\\

CDMHCC & CDM HCC& questioned concepts
\\

HCC & Hierarchical Clustering Cosmology& questioned concept
\\

HGC & Hickson compact Galaxy Cluster& Stephan's Quintet (SQ=HGC 92)
\\

HGT & Hydro-Gravitational Theory& modifies Jeans 1902
\\

ISM &Inter-Stellar Medium& mostly PFPs and gas from PFPs
\\

JPP&Jovian PFP Planet&planet formed by PFP accretion
\\

NBDM & Non-Baryonic Dark Matter&possibly neutrinos
  \\

OGC & Old Globular star Cluster& PGC forms stars at $t
\approx 10^{6}$ yr
\\

PFP&Primordial Fog Particle&planet-mass protogalaxy fragment
\\

PGC & Proto-Globular star Cluster& Jeans-mass protogalaxy fragment
\\

YGC & Young Globular star Cluster& PGC forms stars at $t
\approx$ now
\\


\enddata


\end{deluxetable}

\end{document}